\documentclass[12pt]{iopart}

\usepackage{iopams,graphicx,amssymb}
\eqnobysec

\begin{document}

\title[Effects of random environment on a self-organized  critical system]
{Effects of random environment on a self-organized \\ critical
system: Renormalization group analysis of \\ a continuous model}

\author{N. V. Antonov and P. I. Kakin}

\address{Department of Theoretical Physics, St.~Petersburg University,
Uljanovskaja 1, St.~Petersburg, Petrodvorez, 198504 Russia}

\ead{n.antonov@spbu.ru, p.kakin@spbu.ru}

\begin{abstract}
We study effects of random fluid motion on a system in a self-organized
critical state. The latter is described by the continuous stochastic model,
proposed by Hwa and Kardar [{\it Phys. Rev. Lett.} {\bf 62}: 1813 (1989)].
The advecting velocity field is Gaussian, not correlated in time, with
the pair correlation function of the form
$\propto \delta(t-t') / k_{\bot}^{d-1+\xi}$, where
$k_{\bot}=|{\bf k}_{\bot}|$ and ${\bf k}_{\bot}$ is the component of the
wave vector, perpendicular to a certain preferred direction -- the
$d$-dimensional generalization of the ensemble introduced
by Avellaneda and Majda [{\it Commun. Math. Phys.} {\bf 131}: 381 (1990)].
Using the field theoretic renormalization group we show that, depending on
the relation between the exponent $\xi$ and the spatial dimension $d$, the
system reveals different types of large-scale, long-time scaling behaviour,
associated with the three possible fixed points of the renormalization group
equations. They correspond to ordinary diffusion, to passively advected scalar
field (the nonlinearity of the Hwa--Kardar model is irrelevant) and to the
``pure'' Hwa--Kardar model (the advection is irrelevant). For the special
choice $\xi=2(4-d)/3$ both the nonlinearity and the advection are important.
The corresponding critical exponents are found exactly for all these cases.
\end{abstract}

\pacs{05.10.Cc, 05.70.Fh}

\newpage

\section{Introduction} \label{sec:Intro}

Numerous physical systems reveal self-similar (scaling) behaviour over the
extended ranges of spatial or temporal scales with highly universal
exponents. The most prominent example is provided by the critical behaviour
of various equilibrium systems near their second-order phase transition
points; see, e.g., \cite{Zinn,Book3} and references therein. An essentially
different example is given by the phenomenon of self-organized criticality
(SOC), typical of open nonequilibrium systems with dissipative transport;
see \cite{BTW} and references therein.
In contrast to equilibrium systems, they do not include a tuning parameter
(like the temperature for the second-order transitions) and evolve to the
critical state as a result of their intrinsic dynamics.
Self-organized critical states are believed to be ubiquitous in the Nature
and are observed also in biological, ecological and social systems
\cite{Bak}.

Investigation of the effects of various kinds of deterministic or chaotic
flows on the behaviour
of critical systems (like liquid crystals or binary mixtures near their
consolution points) has shown that the flow can destroy the usual critical
behaviour: it can change to the mean-field behaviour or, under some
conditions, to a more complex behaviour described by new non-equilibrium
universality classes \cite{Satten}--\cite{AIK}.

In this paper, we study effects of turbulent advection on a system in a
self-organized critical state by means of the field theoretic
renormalization group (RG). As a rule, SOC is studied on the base of
discrete models with discrete time steps (like cellular automata).
We employ the continuous model proposed in \cite{HK}, described by the
stochastic equation for a smoothed (coarse-grained) height field
$h(x)=h(t,{\bf x})$ of a certain profile (sand pile or a landscape). Some
modifications of the model \cite{HK} were proposed in \cite{Tadic,Erosia}
and the effects of a flow on the erosion of landscapes were discussed in
\cite{Erosia}, but the velocity field as an independent dynamical variable
was not introduced there.

In this paper, we explicitly introduce coupling with the velocity field.
It is modelled by the strongly anisotropic Gaussian ensemble, with vanishing
correlation time and prescribed power-like pair correlation function,
the $d$-dimensional generalization of the ensemble introduced and studied
in \cite{AM} and, at the same time, the anisotropic modification of the
popular Kraichnan's rapid-change model; see \cite{FGV} for the review and
the references. It is natural here to consider anisotropic flow, because the
original model \cite{HK} already involves intrinsic anisotropy, related
to the overall tilt of the landscape or the sand pile surface.

\section{The model. Field theoretic formulation and renormalization}
\label{sec:QFT}

Let ${\bf n}$ be a unit constant vector that determines a certain preferred
direction. Then any vector can be decomposed into the components
perpendicular and parallel to ${\bf n}$, for example,
${\bf x} = {\bf x}_{\bot} + {\bf n} x_{\parallel}$ with
${\bf x}_{\bot} \cdot {\bf n} =0$. Let
$\partial_{i} = \partial/ \partial {x_i}$ with $i=1\dots d$
be the derivative in the full $d$-dimensional ${\bf x}$ space,
$\partial_{\bot}=\partial/ \partial {x_{\bot i}}$ with $i=1\dots d-1$
is the derivative in the subspace orthogonal to ${\bf n}$, and
$\partial_{\parallel} = {\bf n}  \cdot \partial$.

The stochastic differential equation for the height field
$h(x)=h(t,{\bf x})$ is taken in the form \cite{HK}:
\begin{equation}
\partial_{t} h= \nu_{\bot 0}\, \partial_{\bot}^{2} h + \nu_{\parallel 0}\,
\partial_{\parallel}^{2} h -
\partial_{\parallel} h^{2}/2 + f,
\label{eq1}
\end{equation}
where $\partial_{t} = \partial/ \partial {t}$, $\nu_{\bot 0}$ and
$\nu_{\parallel 0}$ are the bare
diffusivity coefficients and $f(x)$ is a Gaussian random
noise with zero mean and the pair correlation function
\begin{equation}
\langle f(x)f(x') \rangle = 2 D_{0}
\delta(t-t')\, \delta^{(d)}({\bf x}-{\bf x}') .
\label{forceD}
\end{equation}
Coupling with the velocity field $v_{i}(x)$ is introduced by the replacement
$
\partial_{t} \to \nabla_{t} = \partial_{t} + v_{i} \partial_{i},
$
where $\nabla_{t}$ is the Lagrangean (Galilean covariant) derivative.

The velocity field will be taken in the form
$
{\bf v} = {\bf n} v(t, {\bf x}_{\bot}),
$
where $v(t, {\bf x}_{\bot})$ is a scalar function
independent of $x_{\parallel}$. Then the incompressibility condition
is automatically satisfied:
\begin{equation}
\partial_{i} v_{i} = \partial_{\parallel} v(t, {\bf x}_{\bot}) = 0.
\label{inko}
\end{equation}
For $v(t, {\bf x}_{\bot})$ we assume a Gaussian
distribution with zero mean and the pair correlation function
\begin{eqnarray}
\langle v(t,{\bf x}_{\bot}) v(t',{\bf x}_{\bot}') \rangle = \delta(t-t')
\int
\frac{d {\bf k}}{(2\pi)^{d}} \, \exp \left\{ {\rm i}
{\bf k}\cdot ({\bf x}-{\bf x}') \right\}
D_{v} (k)= \nonumber \\
= \delta(t-t') \int \frac{d {\bf k}_{\bot}}{(2\pi)^{d-1}} \, \exp
\left\{ {\rm i} {\bf k}_{\bot}\cdot ({\bf x}_{\bot}-{\bf x}'_{\bot}) \right\}
\widetilde D_{v} (k_{\bot}),
\label{veloc1}
\end{eqnarray}
with $k_{\bot}=|{\bf k}_{\bot}|$ and the scalar coefficient functions
of the form
\begin{equation}
D_{v} (k)= 2\pi \delta(k_{\parallel}) \, \widetilde D_{v} (k_{\bot}) ,
\quad \widetilde D_{v} (k_{\bot}) = B_{0}\, k_{\bot}^{-d+1-\xi}.
\label{veloc2}
\end{equation}
Here $B_{0}>0$ is a constant amplitude factor and $\xi$ is an arbitrary
exponent, which (along with the conventional $\varepsilon=4-d$) will play
the role of a formal RG expansion parameter. The infrared (IR)
regularization in (\ref{veloc1}) is provided by the cutoff $k_{\bot}>m$.
[Its precise form is unimportant; sharp cutoff is merely the most convenient
choice from the calculational viewpoints.]

According to the general statement (see, e.g, the monographs \cite{Zinn,Book3}
and the references therein), our stochastic problem is equivalent to the
field theoretic model of the extended set of fields $\Phi = \{ h', h,
{\bf v} \}$ with action functional
\begin{equation}
{\cal S}(\Phi)=h'D_0 h'+h'\{-\partial_{t}h-v\partial_{\parallel}h+
\nu_{\bot 0} \partial_{\bot}^{2} h +
\nu_{\parallel 0} \partial_{\parallel}^{2} h -
\partial_{\parallel} h^{2}/2\}+ {\cal S}_{\boldsymbol{v}} .
\label{action}
\end{equation}
All the required integrations over $x=\{t,{\bf x}\}$ and
summations over the vector indices are implied.
The last term in (\ref{action}) corresponds to the Gaussian averaging over
${\bf v}$ with correlator (\ref{veloc1}) and has the form
\begin{equation}
{\cal S}_{\boldsymbol{v}}= \frac{1}{2}\,
\int dt \int d{\bf x}_{\bot} d{\bf x}_{\bot}' v(t,{\bf x}_{\bot})
\widetilde D^{-1}_{v} ({\bf x}_{\bot}-{\bf x}'_{\bot}) v(t,{\bf x}_{\bot}'),
\label{Sv}
\end{equation}
where
\begin{equation}
\widetilde D^{-1}_{v} ({\bf r}_{\bot}) \propto D_{0}^{-1} \,
r_{\bot}^{2(1-d)-\xi}
\label{Dv}
\end{equation}
is the kernel of the inverse linear operation $D^{-1}_{v}$ for the
correlation function $D_{v}$ in (\ref{veloc2}).

This allows one to apply
the field theoretic renormalization theory and renormalization group to
our stochastic problem. The model (\ref{action}) corresponds to a standard
Feynman diagrammatic technique with three bare propagators:
$\langle {\bf v}{\bf v} \rangle_{0}$, given by (\ref{veloc1}),
(\ref{veloc2}), and the propagators of the scalar fields
(in the frequency--momentum representation):
\begin{equation}
\langle hh' \rangle_{0} = \langle h'h
\rangle_{0}^{*}
= \left\{-{\rm i} \omega+ \varepsilon(k)
\right\}^{-1}, \quad
\langle hh \rangle_{0} =  2 D_{0} \,
\left\{ \omega^{2} + \varepsilon^{2}(k) \right\}^{-1},
\label{lines3}
\end{equation}
where $\varepsilon(k)=\nu_{\parallel 0}^2 k_{\parallel}^2+\nu_{\bot 0}^2
k_{\bot}^2$. The propagator $\langle h'h'\rangle_{0}$ vanishes identically
for any field theory of the type (\ref{action}). The model also involves
two types of vertices corresponding to the interaction terms
$-h'\partial_{\parallel} h^{2}/2$ and $-h'(v\partial_{\parallel})h$.
The corresponding coupling constants
$g_{0}$ and $w_{0}$ are defined  by the relations
\begin{equation}
D_{0} = \nu_{\parallel 0}^{3/2}\nu_{\bot 0}^{3/2} g_{0}, \quad
B_{0} = w_{0} \nu_{\parallel 0},
\label{D0}
\end{equation}
so that by dimension $g_{0} \sim \ell^{-\varepsilon}$ and
$w_{0} \sim \ell^{-\xi}$,
where is $\ell$ has the order of the smallest length scale in our problem.

The analysis of canonical dimensions (see, e.g., \cite{Zinn,Book3})
along with symmetry considerations shows that our model
is multiplicatively renormalizable with the only
independent renormalization constant $Z_{\nu_{\parallel}}$:
all the renormalization constants are equal to $1$. In the leading
(one-loop) approximation $Z_{\nu_{\parallel}}$ has the form
$Z_{\nu_{\parallel}}=1-a{g}/{\varepsilon}-b{w}/{\xi}$
(the use of the minimal subtraction scheme is implied).
Here $g$ and $w$ are renormalized analogues of the bare parameters
in (\ref{D0}), which play the role of the coupling constants (dimensionless
expansion parameters) in renormalized perturbation theory, and $a,b$
are numerical coefficients. Their precise values are unimportant
(they can simply be absorbed by the redefinition of $g$ and $w$),
but it is important for the following that they are positive:
$a,b>0$.

To proceed, we have to find the RG functions: the anomalous dimensions
and the $\beta$ functions, which appear as coefficients in the
differential RG equations; detailed discussion can be found in
\cite{Zinn,Book3}. The anomalous dimension $\gamma_{F}$ for a given
renormalization constant $Z_{F}$ is defined as
\begin{equation}
\gamma_{F}\equiv \widetilde{\cal D}_{\mu} \ln Z_{F}
\quad {\rm for\ any\ quantity} \ F,
\label{RGF1}
\end{equation}
where $\widetilde{\cal D}_{\mu}$ is the differential operation
$\mu\partial_{\mu}$ for fixed $e_{0}$, the latter being the full set of
bare parameters $e_{0}=\{\nu_{\parallel 0}, \nu_{\bot 0}, w_{0}, g_{0} \}$.
The $\beta$ functions for the two dimensionless couplings $g$ and $w$ are
\begin{equation}
\beta_{g} \equiv \widetilde {\cal D}_{\mu} g = g\,[-\varepsilon-\gamma_{g}],
\quad
\beta_{w} \equiv \widetilde {\cal D}_{\mu} w = w\,[-\xi-\gamma_{w}].
\label{betagw}
\end{equation}

Taking into account the relations between the bare parameters (\ref{D0}),
absence of renormalization of $D_{0}$, $B_{0}$
and the explicit form of the constant $Z_{\nu_{\parallel}}$, we obtain:
\begin{equation}
\beta_{g} = g [-\varepsilon+ 3\gamma_{\nu_{\parallel}}/2], \quad
\beta_{w} = w [-\xi+ \gamma_{\nu_{\parallel}}], \quad
\gamma_{\nu_{\parallel}} = ag + bw + \dots
\label{gammas}
\end{equation}
where the ellypsis stands for the higher-order corrections
in $g$ and $w$.

\section{Fixed points and scaling regimes} \label{sec:FPS}

It is well known that possible scaling regimes of a renormalizable model are
associated with IR attractive fixed points of the
corresponding RG equations. In the present case, the coordinates
$g_{*}$, $w_{*}$ of the fixed points are found from the equations
\begin{equation}
\beta_{g} (g_{*},w_{*}) = 0, \quad \beta_{w} (g_{*},w_{*})=0 ,
\label{points}
\end{equation}
with the $\beta$ functions given in (\ref{gammas}).
The type of a fixed point is determined by the matrix
$
\Omega=\{\Omega_{ij}=\partial\beta_{i}/\partial g_{j}\}|_{g_{*},w_{*}},
$
where $\beta_{i}$ is the full set of the $\beta$ functions and
$g_{j}= \{g,w\}$ is the full set of coupling constants. For
IR attractive points
the matrix $\Omega$ is positive, that is, the real parts of
all its eigenvalues $\Omega_{i}$ are positive.

The $\beta$ functions (\ref{gammas}) involve the same anomalous dimension
$\gamma_{\nu_{\parallel}}$ and thus satisfy the exact relation
$w\beta_{g}-3g\beta_{w}/2 =gw (-\varepsilon+3\xi/2)$. In turn, this means
that for general $\varepsilon$ and $\xi$ the equations (\ref{points})
can be satisfied only if at least one of the coordinates
$g_{*}$, $w_{*}$ vanishes identically. There are three such fixed points:

(1) Gaussian (free) fixed point with $g_{*}=w_{*}=0$ and
$\gamma_{\nu_{\parallel}}^{*}\equiv\gamma_{\nu_{\parallel}}(g_{*}=w_{*}=0)=0$.
The eigenvalues of the matrix $\Omega$ are equal to its diagonal
elements: $\Omega_{g}=\partial_{g}\beta_{g}= -\varepsilon$,
$\Omega_{w}=\partial_{w}\beta_{w}=-\xi$. Thus, the point is attractive
for $\varepsilon$, $\xi<0$.

(2) The point with $g_{*}=0$, $w_{*}=\xi/b$ and
$\gamma_{\nu_{\parallel}}^{*}=\xi$. The latter expression is exact (no
corrections of order $\xi^2$ and higher) owing to the exact relation
between $\beta_{w}$ and $\gamma_{\nu_{\parallel}}$ in (\ref{gammas}).
Furthermore, the derivative
$\partial_{w}\beta_{g} = 3g \partial_{w} \gamma_{\nu_{\parallel}}/2$
vanishes for $g_{*}=0$, the matrix $\Omega$ is block triangular and  its
eigenvalues coincide with the diagonal elements:
$\Omega_{g}=-\varepsilon+3\xi/2$ (exact result) and
$\Omega_{w}=\xi$ (with possible corrections of order $\xi^2$ and higher).
We conclude that this point is attractive for $\xi>2\varepsilon/3$, $\xi>0$.

From the physics viewpoints, in this critical regime the nonlinear
term in the equation (\ref{eq1}) does not affect the leading terms of the
IR asymptotic behaviour (it is IR irrelevant in the sense of Wilson).

(3) The point with $w_{*}=0$, $g_{*}=\varepsilon/a$ and
$\gamma_{\nu_{\parallel}}^{*}=2\varepsilon/3$.
The latter expression is exact (no
corrections of order $\varepsilon^2$ and higher) owing to the exact relation
between $\beta_{g}$ and $\gamma_{\nu_{\parallel}}$ in (\ref{gammas}).
Furthermore, the derivative
$\partial_{g}\beta_{w} =w \partial_{g} \gamma_{\nu_{\parallel}}$
vanishes for $w_{*}=0$, the matrix $\Omega$ is block triangular and its
eigenvalues again coincide with the diagonal elements:
$\Omega_{w}=-\xi+2\varepsilon/3$ (exact result) and
$\Omega_{w}=\varepsilon$ (with corrections of order $\varepsilon^2$ and
higher).
We conclude that this point is attractive for $\xi<2\varepsilon/3$,
$\varepsilon>0$. In this critical regime, the advection is irrelevant
and the IR behaviour of the model coincides with that of the original
Hwa--Kardar model.

It is also worth noting, that for all fixed points (1)--(3) in the
regions of their IR attraction, their
coordinates lie in the physical region $g_{*}\ge0$, $w_{*}\ge0$.

For the special choice $\xi=2\varepsilon/3$ the functions (\ref{gammas})
become proportional, and equations (\ref{points}) only impose the
restriction $ag_{*}+bw_{*}=\xi$; that is, we have a line of fixed points
in the $g$--$w$ plane. However, the exact value
$\gamma_{\nu_{\parallel}}^{*} = \xi=2\varepsilon/3$ is the same for all
points on that line. The direct calculation of the elements of the
matrix $\Omega$ shows that one of the eigenvalues vanishes (this
exact result is a consequence of the degeneracy of the fixed point)
while the other, equal to $3ag_{*}/2+bw_{*}$, is positive in the physical
region $g_{*},w_{*}>0$, as well as the exponent $\xi$.

\begin{figure}
\begin{center}
\includegraphics[width=7cm]{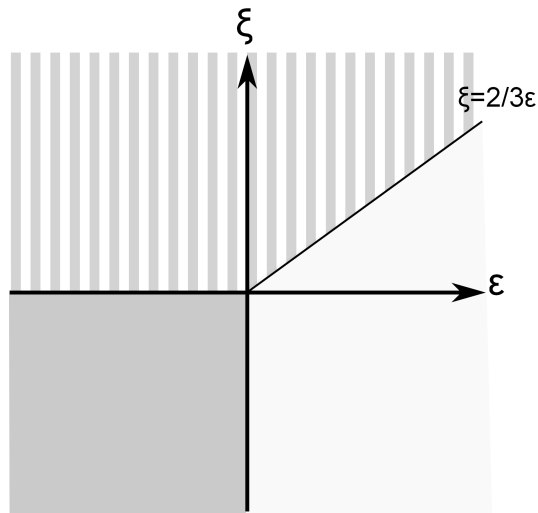}
\caption{\label{fig:pattern} Regions of attraction of the fixed points
in the model (\protect\ref{action}). It should be noted that all the
boundaries of those regions are straight rays; there are neither gaps
nor overlaps between the different regions. These results are exact (not
affected by the higher orders $\xi^2$, $\varepsilon^{2}$  and so on.)}
\end{center}
\end{figure}

In figure~\ref{fig:pattern} we show the regions of IR attraction for all the
fixed points in the $\varepsilon$--$\xi$ plane, that is, the regions where
the eigenvalues of the matrix $\Omega$ are both positive.

\section{Critical scaling and critical dimensions} \label{sec:DimeNS}

Existence of IR attractive fixed points of the RG equations leads to
appearance of the scaling behaviour of the Green functions in the IR range.
In dynamical models with two momentum scales, like our present model, the
critical dimension $\Delta_{F}$ of an IR relevant
quantity $F$ (a field or a parameter) is given by the relation
\begin{equation}
\Delta_{F} = d_{F}^{\bot} +  d_{F}^{\parallel} \Delta_{\parallel} +
d_{F}^{\omega} \Delta_{\omega} + \gamma_{F}^{*}.
\label{dim}
\end{equation}
Here $d_{F}$ are the canonical dimensions (see, e.g., \cite{Book3}),
$\gamma_{F}^{*}$ is the fixed-point value of the corresponding anomalous
dimension,
\begin{equation}
\Delta_{\parallel} = 1 + \gamma_{\nu_{\parallel}}^{*}/2, \quad
\Delta_{\omega}=2 -\gamma_{\nu_{\bot}}^{*}
\label{dimV}
\end{equation}
are the critical dimensions of the momentum in the direction parallel
to ${\bf n}$ and of the
frequency, respectively, with the normalization $\Delta_{\bot}=1$
for the momentum in perpendicular subspace.
Detailed derivation and discussion of the relations of the type
(\ref{dim}), (\ref{dimV}) can be found, e.g., in ref.~\cite{Alexa}.

In our model these relations simplify owing to the absence of renormalization
of all the fields (so that $\gamma_{F}^{*}=0$) and of the parameter
$\nu_{\bot 0}$ (so that $\gamma_{\nu_{\bot}}^{*}=0$); see sec.~\ref{sec:QFT}.
What is more, $\gamma_{\nu_{\parallel}}^{*}$ is known exactly for all the
fixed points. This gives the following exact expressions:
\begin{equation}
\Delta_{h} = 1, \quad \Delta_{h'} = d = 4-\varepsilon, \quad
\Delta_{\omega} =2, \quad \Delta_{\parallel} =1
\label{dimG}
\end{equation}
for the Gaussian fixed point (1),
\begin{equation}
\Delta_{h} = 1 - \xi/2, \quad \Delta_{h'} = d+\xi = 4-\varepsilon+\xi, \quad
\Delta_{\omega} =2, \ \Delta_{\parallel} =1+\xi/2
\label{dimA}
\end{equation}
for the fixed point (2) and the line of the fixed points, and
\begin{equation}
\Delta_{h} = 1 - \varepsilon/3, \quad
\Delta_{h'} = d+2\varepsilon/3 = 4-\varepsilon/3, \quad
\Delta_{\omega} =2, \ \Delta_{\parallel} =1+\varepsilon/3
\label{dimS}
\end{equation}
for the point (3).

In particular, for the pair correlation function of the main field this
gives:
\begin{equation}
\langle h(t,{\bf x})\, h(0,{\bf 0}) \rangle \simeq
r_{\bot}^{-2\Delta_{h}}\, {\cal F} \left(t/ r_{\bot}^{\Delta_{\omega}},
r_{\parallel}/ r_{\bot}^{\Delta_{\parallel}} \right),
\label{dimS1}
\end{equation}
where $r_{\bot}=|{\bf x}_{\bot}|$, $r_{\parallel}=x_{\parallel}$ and
${\cal F}$ is a certain scaling function  of critically dimensionless
arguments.

It remains to note that for the fixed point (3) our results agree, up to
the notation, with those obtained in \cite{HK} (one has to identify
$z=\Delta_{\omega}/\Delta_{\parallel}$, $\zeta=1/\Delta_{\parallel}$,
$\chi = -\Delta_{h}/\Delta_{\parallel}$).

\section{Conclusion} \label{sec:Conc}

We studied effects of fluid motion, including turbulent mixing, on a
system in the state of self-organized criticality. The latter was modelled
by a stochastic differential equation for a height field
(\ref{eq1}), (\ref{forceD}),
while the velocity of the fluid was described by a random Gaussian ensemble
(\ref{veloc1}). Both the equation and the velocity ensemble are strongly
anisotropic.

The full model can be reformulated as a renormalizable field theoretic model
with a single independent renormalization constant. The corresponding RG
equations possess several IR attractive fixed points, corresponding to
different possible types of critical behaviour.
They correspond to ordinary diffusion, passively advected scalar
field and to the original SOC model without the mixing.
Their regions of IR attraction in the plane of the model parameters
$d$ and $\xi$ and the critical dimensions of the basic fields and parameters
are found exactly. For the special choice $\xi=2(4-d)/3$ an intermediate
regime arises, where both the nonlinearity and the advection are important.

In the following, it would be interesting to study other models of SOC
and to employ more realistic models for the turbulent velocity field, e.g.,
models with finite correlation time and non-Gaussianity. This work is in
progress.

\newpage

\section*{Acknowledgments}

The authors thank Loran Adzhemyan and Michal Hnatich for discussion.
The authors also thank the Organizers of the conference ``Mathematical
Modeling and Computational Physics 2015''
(Star\'a Lesn\'a in High Tatra Mountains, 13--17 July 2015)
for the possibility to present the results of the present study.

The work was supported by the Saint Petersburg State University within the
research grant 11.38.185.2014.

%\newpage

\end{document}